# A survey of blockchain frameworks and applications

Bruno Tavares[1[0000-0001-8307-3923]], Filipe Figueiredo Correia[12[0000-0002-6653-1598]], André Restivo[1[0000-0002-1328-3391]], João Pascoal Faria[12[0000-0003-3825-3954]], Ademar Aguiar[12[0000-0002-4046-4729]]

[1] Department of Informatic Engineering, Faculty of Engineering, University of Porto, Portugal
[2] INESC TEC, Porto, Portugal

{up201700372,filipe.correia,arestivo,jpf,ademar.aguiar}@fe.up.pt

**Abstract.** The applications of the blockchain technology are still being discovered. When a new potential disruptive technology emerges, there is a tendency to try to solve every problem with that technology. However, it is still necessary to determine what approach is the best for each type of application. To find how distributed ledgers solve existing problems, this study looks for blockchain frameworks in the academic world. Identifying the existing frameworks can demonstrate where the interest in the technology exists and where it can be missing. This study encountered several blockchain frameworks in development. However, there are few references to operational needs, testing, and deploy of the technology. With the widespread use of the technology, either integrating with pre-existing solutions, replacing legacy systems, or new implementations, the need for testing, deploying, exploration, and maintenance is expected to intensify.

## 1 Introduction

Research groups are looking for new applications for the blockchain technology. Although this technology has achieved popularity in the implementation of decentralized crypto currencies, the technology offers many capabilities that can be advantageous in different applications. This study aims to find what research works exist for blockchain frameworks that assist in the development of blockchains or blockchain-related applications. As for new technologies, sometimes the academic world and the industrial world work at different paces, so it is also important to find out where the innovation is coming from, to find out what approaches are still missing [27].

This survey intends to find out what is the types of scientific contribution each paper contains. The analysis of research works about blockchain frameworks over the past few years, and the analysis of the metadata of academic literature can occasionally reveal important information about the technology.

To understand the impact blockchain can create, it is necessary to understand what does blockchain provide. Understanding from a high-level perspective how the technology works and what capabilities it possesses can help the analyses of the literature.

The second section of this paper presents the blockchain technology from a high-level perspective and some blockchain applications. The third section presents metadata information about blockchain-related academic literature. The forth section contains the scientific contribution for each paper and contains the analysis of the research works. In the fifth and last section the author presents his conclusions.

## 2 Blockchain

### 2.1 Blockchain concept

A blockchain is a series of blocks connected to each other, where the last block connects to the previous one and so on. This distributed ledger leverages cryptographic techniques, that can time-stamp information in a system that is immune to fraud or central control if the processing power of the blockchain is distributed by different nodes. Each block holds information; when a node creates a block it will try to add the block to the blockchain. The other nodes need to verify each block. When the validation of the block is complete, the nodes will accept the new block, extending the blockchain. The algorithms, used to handle the verification and consensus between multiples entities, can be different for each blockchain implementation.

**Fig. 1** shows how the blocks connect to each other in a blockchain [29].

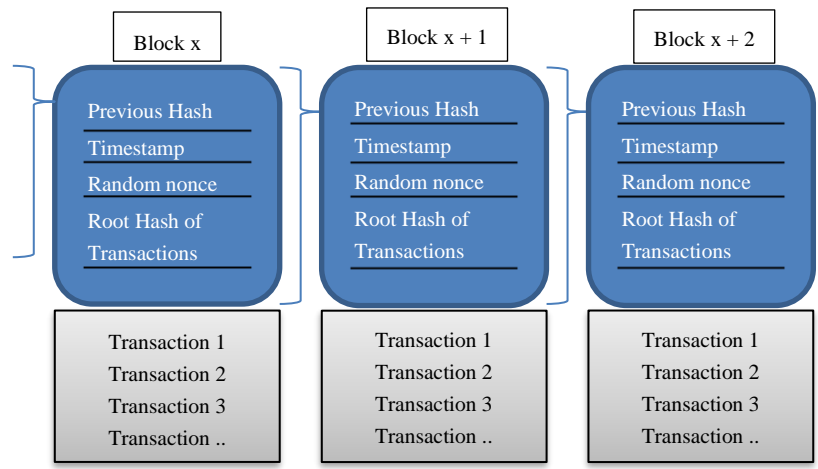

**Fig. 1.** Blockchain scheme

The top part of the block is the header and the lower part is the body of the block. The body contains the transactions of that block. The block header contains the previous hash; the previous hash is a hash of the header of the previous block. Adding the hash of the previous block to a new block creates a chain. The header also contains the timestamp, a random nonce, and the root hash of the transactions in the body. The random nonce is an arbitrary number used once. The root hash of transactions is the top hash of a Merkle tree demonstrated in **Fig. 2** [31].

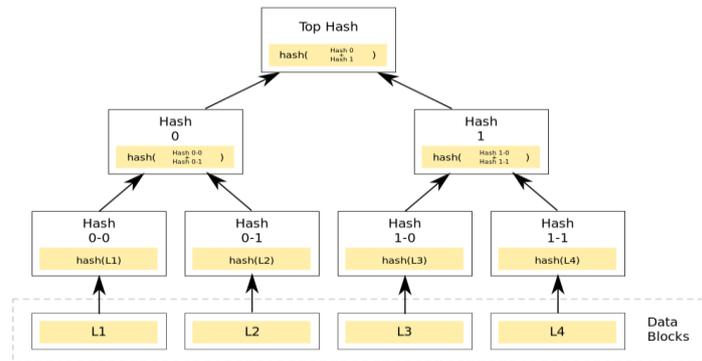

**Fig. 2.** Merkle tree for root hash of transactions

## 2.2 Blockchain capabilities

The blockchain technology offers different capabilities, but the main benefits that the technology offers are the ability to mitigate the need for a trusted third party. When dealing with other companies or individuals, some level of trust is necessary. In today's world people trust the good name of the company or third-parties' certifications. Blockchain offers an alternative, moving the trust needed from an entity to the technology itself. In summary, blockchain is a shared ledger that provides a consensus mechanism with a certain degree of anonymity and privacy in an untrusted network. Blockchain technology can also be the ground for smart contracts that allow transactions with different types of validations where the verification occurs in a transparent system. In the end, the transactions are immutable and it is not possible to tamper the record [27]. The main benefits the technology has to offer are the following [13]:

- **Trustless –** The entities are in control of the information and two entities can make an exchange without the need of a trusted third party.
- **Integrity –** The execution of transactions is exactly as the protocol commands.
- **Transparency –** Modifications to public blockchains are public and can be analyzed and accessible to all entities creating transparency.
- **Quality data** – The data in the blockchain is complete, consistent and immutable.
- **Reliability –** Blockchain uses a decentralized network; this eliminates the central point of failure problem and improves survival from attacks.

## 2.3 Blockchain challenges

Blockchain is a technology that contains several challenges. Transaction speed, limit of information each block can contain, and the verification processes need validation in different scenarios to make the technology widely accepted. Depending on the consensus algorithm used a distributed cryptographic ledger can also consume a large amount of energy. Regarding control, security, and privacy, it is necessary further research, because, although all data can be encrypted, the data is shared, and if the encryption protocol is broken, then the information will be exposed [27].

For the technology to be able to replace existent systems, it is necessary to overcome, at least, the following problems: integration issues, cultural adoption, and uncertain regulation. The integration concerns exist because the blockchain application solutions requires significant changes and, most of the times, complete replacement of current systems. It is necessary for the entities to organize and define a strategy for a technology evolution. The cultural adoption is dependent on the robustness and strength of current applications of the blockchain. Most people see the technology as something new but not as a disruptive technology, and there is some skepticism around the true capabilities of blockchain [27]. For a technology shift, and a widespread adoption of the blockchain technology, regulation is necessary. Although blockchain promises to solve the trusted third party issue, it is still required that the regulation and authorities enforce by legal means any dispute that may arise [12].

### 2.4 Blockchain applications

The blockchain technology has been used to put proof of existence in legal documents, health records, supply chain, IoT E-business, energy market, E-Governance, decentralized registry, stock exchanges and smart cities [27]. The economic, legal and political systems deal with contracts, transactions, and records. They protect assets and set organizational restrictions. They create and authenticate information. This information allows interactions among countries, societies, business, and individuals. Any system or organization that wants to share information with transparency and safe immutability between different parties, can leverage the capabilities offered by the blockchain technology [12].

There are several applications of the technology in the finance world. Fintech companies are starting to explore cryptocurrencies and exchange markets. In addition, to explore the blockchain technology for business-to-business transactions, prospective benefits include cyber risk reduction, counterparty risk, and increase of transparency [13][8]. In healthcare the information from patients and medical research data need to be shared, in a way where sensitive personal information is not revealed but also cannot be tampered [10]. The technology can also be applied in organizations responsible for: civil registries, deaths, marriages, land registries, etc. [28].

The energy market is exploring the blockchain technology to reduce the cost of existent centralized solutions, where maintenance and management is too high. It is demanding to support the collection, transaction, storage and analysis of substantial data in the energy market and a peer-to-peer solution can help to mitigate this problem [7].

Internet of Things (IoT) is a new phenomenon predisposed to deliver several services and applications. Developing technologies at the same time can promote work synergies, methods, and frameworks, capable of enabling the best features from each technology [12][23].

## 3 Research works about blockchain

The meta information that exists about a technology can indicate the progress and interest of a subject in the academic community. **Fig. 3** illustrates the number of papers related with the blockchain keyword found in Scopus when this survey was made.

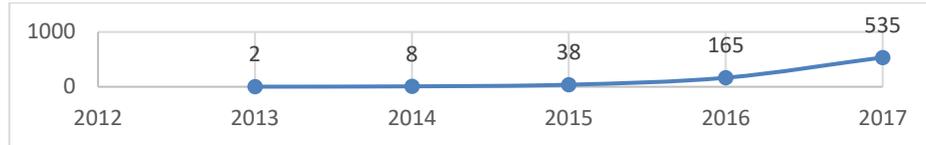

**Fig. 3.** Number of Blockchain related papers in Scopus

Another interesting metric is the one presented in **Fig. 4**. There are still unknown applications of blockchain. However, a search of the related papers by subject area can reveal that many different areas are studying a way to apply the technology.

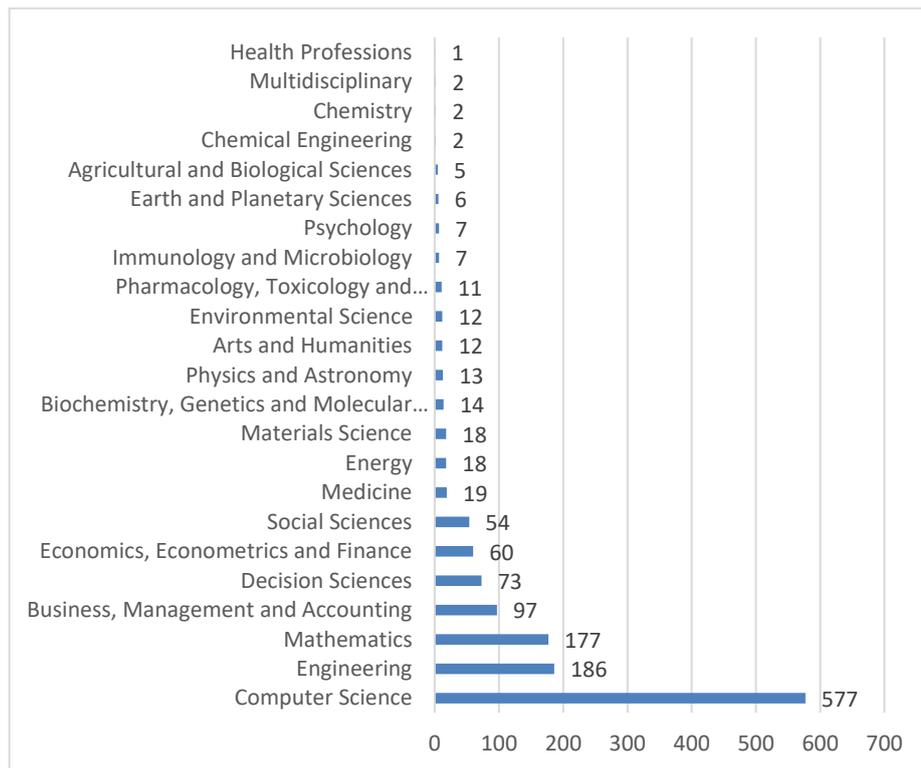

**Fig. 4.** Blockchain related papers by subject area in Scopus

The interest in blockchain is increasing exponentially in the last couple of years. However, the academic interest is related to a commercial interest. Many businesses are trying to leverage the blockchain capabilities to solve real-world problems [21][19].

In **Fig. 4** chart, the same paper may relate to more than one subject area. Nevertheless, this information indicates that there are many potential applications of the technology in very distinct fields.

**Fig. 5** classifies blockchain related papers by document type. This information does not vary from subject to subject. Usually there are more conference papers and articles indexed by Scopus.

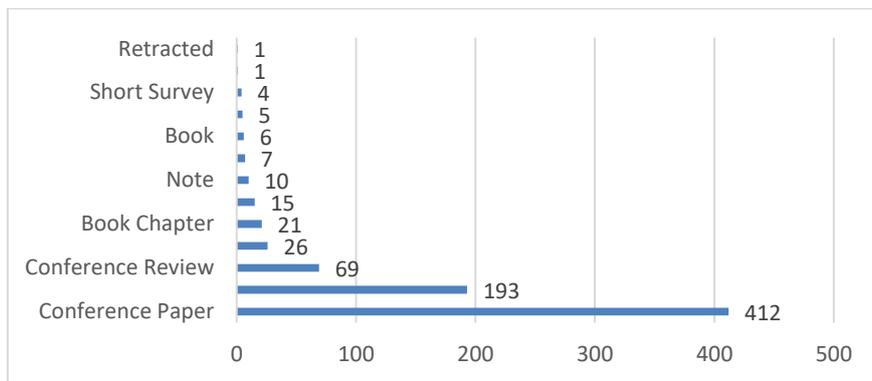

**Fig. 5.** Blockchain related papers by document type in Scopus

**Fig. 6** shows the number of citations that the blockchain papers have received. This information shows that the more relevant papers are a couple of years old.

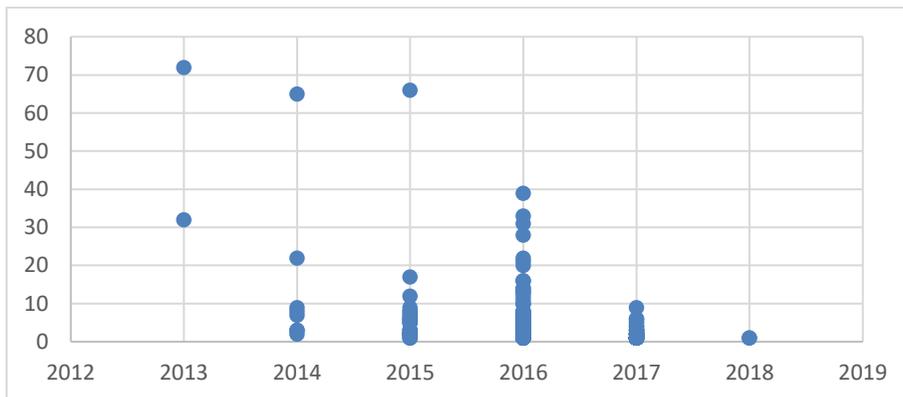

**Fig. 6.** Blockchain related papers by citations in Scopus

# 4 Research works about blockchain frameworks

This research is based in a Scopus search with the keywords: blockchain and frameworks; the search query was also limited to conference papers and articles. This analysis focuses on frameworks created for blockchain technology.

**Table 1.** Research works about blockchain frameworks

| Author | Framework | Area | Scientific Contribution |
|---|---|---|---|
| Cha et al. (2018)[5] | Smart Contract-based Investigation Report Management framework for smartphone applications security (SCIRM) | Certification | Architecture Interfaces |
| Allen et al. (2017)[1] | Institutional Possibility Frontier framework | E-Govern | Algorithm |
| Ateniese et al. (2017)[2] | New framework that makes it possible to re-write or compress the content | No specific area | Algorithm |
| Baracaldo et al. (2017)[3] | Framework for cryptographic provenance data protection and access control | IoT | Architecture |
| Chen et al. (2017)[6] | Theoretical framework for evaluating a PoET based blockchain system | Not specific to any area | Analysis |
| Cheng et al. (2017)[7] | Transaction framework is based on the blockchain technology in the distributed electricity market | Energy Market | Analysis Algorithm |
| Dinh et al. (2017)[9] | BLOCKBENCH, the first evaluation framework for analyzing private blockchains | Fintech | Analysis |
| Dubovitskaya et al. (2017)[10] | Framework for managing and sharing EMR data for cancer patient care | Healthcare | Architecture |
| Fabiano N. (2017)[12] | Global privacy standard framework | IoT | Analysis |
| Geranio M. (2017)[13] | Common framework and a proper management of risks | Fintech | Survey |
| Goyal et al. (2017)[14] | Non-interactive zero-knowledge (NIZK) system | Fintech | Algorithm |
| Hou H. (2017)[17] | Chancheng's project called "The Comprehensive Experimental Area of Big Data in Guangdong Province" | E-Govern | Analysis |
| Kinai et al. (2017)[20] | Blockchain backend built on hyperledger fabric and mobile phone applications for stakeholders to engage with system | Fintech | Architecture |

| Li M.-N. (2017) [22] | Conceptual framework of blockchain | Academic | Analysis |
|---|---|---|---|
| Liu et al. (2017)[23] | Blockchain-based framework for Data Integrity Service | IoT | Architecture |
| Ouyang et al. (2017)[25] | Framework of large consumers direct power trading based on blockchain technology | Energy Market | Architecture |
| Wang et al. (2017)[30] | Blockchain router, which empowers blockchains to connect and communicate cross chains | Communication | Architecture |
| Ye et al. (2017)[34] | Customized reputation system framework (DC-RSF) to evaluate the credibility of cloud service vendors | Cloud | Architecture |
| Yuan et al. (2017) [35] | Parallel blockchain aims to offer the key capabilities including computational experiments and parallel decision-making to blockchains via parallel interactions | Academic | Analysis |
| Hari et al. (2017) [16] | Blockchain based mechanism to secure the Internet BGP and DNS infrastructure | Communication | Algorithm |
| Lemieux V.L. (2017)[28] | General evaluative framework for a risk-based assessment of a specific proposed implementation of Blockchain technology | E-Govern | Architecture |
| Yasin et al. (2017)[33] | Systematic framework for aggregating online identity and reputation information | E-Business | Survey |
| Zhang et al. (2017)[36] | Universal Composability (UC) framework, authenticated data feed system called Town Crier (TC) | E-Business | Architecture |
| Zhang et al. (2017) [37] | Blockchain in Energy Internet were analyzed from three dimensions-function, entity and property | Energy Market | Analysis |

## 5 Conclusions

The blockchain technology is a novel technology that gained considerable momentum in the last couple of years. It is a distributed ledger and a possible disruptive technology; the way blockchain manages information, offers a new way of doing things. Creating new foundations and giving meaning to smart contracts, the application of the technology is still in the beginning. There are several frameworks related with the technology; some frameworks solve problems related with specific areas such as healthcare, IoT, energy market and so on. Other frameworks are developed with the type of blockchain in mind [9].

There are still open challenges and replacing legacy systems with new implementations is a process that takes a lot of time and preparation. Although many problems are known, whenever the technology is applied in new environments, new difficulties will arise. Being able to define a framework concentrated in the lifecycle of a blockchain-based project, can probably identify problems before hand and deliver a tested and reliable solution [17].